\documentclass[fontsize=11pt,twoside=semi,numbers=endperiod]{scrartcl}
\KOMAoptions{headinclude=true,footinclude=false,index=toc}
\usepackage[paperwidth=150mm,paperheight=179.35mm,width=110mm,totalheight=165.35mm,includehead,headheight=5.258mm,headsep=1mm,top=6mm,inner=20mm]{geometry} 

\usepackage{amsmath}
\usepackage{amssymb}

\usepackage{lmodern} 

\usepackage[ansinew]{inputenc}
\usepackage[T1]{fontenc}
\usepackage[english,ngerman]{babel}
\usepackage{textcomp} 
\usepackage[pdftex]{color,graphicx} %
\usepackage[rel]{overpic} 
\usepackage{bbm}
\usepackage{pifont}
\usepackage{cite} 
\usepackage{ellipsis} 
\usepackage{microtype} 
\usepackage{scrpage2}
\usepackage[%
	pdftex,
	colorlinks,citecolor=blue,filecolor=blue,linkcolor=blue,urlcolor=blue,%
	bookmarks=true,%
	bookmarksopen=false,%
	bookmarksnumbered=true,%
	pdfpagemode=UseNone,
    pdfstartview={XYZ null null 1.00},
	pdffitwindow=true,
    pdfview={XYZ null null null},
    pdfpagelayout=SinglePage,
    pdfdisplaydoctitle=true, 
    plainpages=false,
    pdfpagelabels=true,
	pdfkeywords={quantum field theory, zero-point energy, quantum vacuum},
	pdftitle={The zero-point energy of elementary quantum fields},
    pdfauthor={Astrophysical Institute Neunhof}]
	{hyperref}

\setkomafont{disposition}{\rmfamily} 
\addtokomafont{section}{\large\bfseries} 
\addtokomafont{subsection}{\normalsize\bfseries} 
\addtokomafont{pagenumber}{\LARGE}
\addtokomafont{caption}{\small}

\deffootnote[1em]{1em}{1em}{\textsuperscript{\thefootnotemark}\ }

\newcommand{\ggitem}[2][$\ast $]{\makebox[0em][r]{#1\;}{#2}\newline }%
\newenvironment{ggitemize}{\vspace{0.2\baselineskip} \begin{addmargin}[1.2em]{0em}}{\vspace{-0.8\baselineskip} \end{addmargin}}

\newcommand{\ggie}{\mbox{i.\,e.}\ } 
\newcommand{\ggeg}{\mbox{e.\,g.}\ } 
 
\newcommand{\bz}[1]{\nolinebreak\hspace{0em}\nolinebreak{}#1\hspace{0em}}
\newcommand{\al}{\iflanguage{english}{``\nolinebreak\hspace{0em}}{\glqq\nolinebreak\hspace{0.1em}}\nolinebreak}
\newcommand{\ar}{\nolinebreak\iflanguage{english}{\hspace{0em}\nolinebreak{}''\ }{\hspace{-0.45em}\nolinebreak\grqq\thickspace}}
\newcommand{\arp}{\nolinebreak\iflanguage{english}{\hspace{0em}\nolinebreak{}''}{\grqq}}

\newcommand{\gginfty}{\infty} 
\newcommand{\ggginfty}{\infty} 

\newcommand{\ggstackrel}[2][0]{\hspace{0.36em}\hspace{#1em}&=\hspace{-#1em}\hspace{-1.16em}\stackrel{#2}{\phantom{=}}} 
\newcommand{\textstackrel}[2][=]{\raisebox{-1mm}[0mm][0mm]{$\stackrel{#2}{#1}$}} 

\newcommand{\inte }{\int \limits }

\newcommand{\mklammerp}{\mbox{\hspace{.4em}\raisebox{1.1mm}[0mm][0mm]{$-$}\hspace{-1em}\raisebox{-.4mm}[0mm][0mm]{\scalebox{.8}[.55]{$($}\scalebox{.8}{$+$}\scalebox{.8}[.55]{$)$}}\hspace{.2em}}}
\newcommand{\pklammerm}{\mbox{\hspace{.4em}\raisebox{1.1mm}[0mm][0mm]{\scalebox{.8}{$+$}}\hspace{-1em}\raisebox{-.4mm}[0mm][0mm]{\scalebox{.8}[.55]{$($}\raisebox{-.4mm}[0mm][0mm]{$-$}\scalebox{.8}[.55]{)}}\hspace{.2em}}}
\DeclareMathOperator{\dif }{d}
\DeclareMathOperator{\Dif }{D}

\clearscrheadfoot 
\rohead[\pagemark ]{\pagemark}
\lehead[\pagemark ]{\pagemark}
\rehead{ }
\lohead{ }

\begin{document}
\predisplaypenalty=0 
\selectlanguage{english}
\thispagestyle{empty}
\raggedbottom  
\vspace*{3mm}
\centerline{\bfseries\Large The zero-point energy of}
\vspace*{.5mm}
\centerline{\bfseries\Large elementary quantum fields}
\vspace*{2mm}
\centerline{{\bfseries Gerold Gr\"{u}ndler}\,\footnote{\href{mailto:gerold.gruendler@astrophys-neunhof.de}{email:\ gerold.gruendler@astrophys-neunhof.de}}}
\vspace*{1mm}
\centerline{\small Astrophysical Institute Neunhof, N\"{u}rnberg, Germany} 
\vspace*{3mm}
\noindent\parbox{\textwidth}{\small Since 1925, exactly four arguments have been forwarded for the assumption of a diverging (respectively --- after regularization --- very huge) zero\bz{-}point energy of elementary quantum fields. And exactly three arguments have been forwarded against this assumption. In this article, all seven arguments are reviewed and assessed. It turns out that the three \textsc{contra} arguments against the assumption of that zero\bz{-}point energy are overwhelmingly stronger than the four \textsc{pro} arguments.\vspace{1mm}\vspace{1mm}}\vspace*{1mm}

\section{Do elementary quantum fields have a non vanishing zero-point energy?} 
The title question of this section is almost as old as quantum theory. In an article, submitted to the \foreignlanguage{ngerman}{\emph{Zeitschrift für Physik}} in November 1925, Born, Heisenberg, and Jordan\!\cite{Born:qm2} computed the energy eigenvalues $E_n$ of a quantum system of harmonic oscillators with eigen\bz{-}frequencies $\nu _j$\,: 
\begin{align}
E_n=h\cdot\sum _j(n_j+1/2)\nu _j\quad\text{with}\ n_j=0,\,1\,,2\,\dots\label{mfdmnfmngh}
\end{align}
With regard to this result, they remarked (my translation): \al The `zero\bz{-}point energy' $\tfrac{1}{2}h\sum _j\nu _j$ [\dots ] would be infinitely large in particular in the limit of infinitely many degrees of freedom.\ar  The wording \al would be \dots in the limit\ar  reveals, that the authors had substantial doubts whether this result should be taken seriously, or whether it might indicate a flaw of the theory. 

Continuous fields have infinitely many degrees of freedom, while discrete fields have only a finite number of degrees of freedom. The word \al discrete\ar  does not refer to the quanta of the discrete field, but to it's material substrate. The phonon field, for example, is a discrete field, because the atom grid of the molecule or solid, which is the material substrate of the phonon field, is a discrete grid, and consequently the phonon field's wavenumber spectrum is limited to the first Brillouin zone. Elementary quantum fields (like the electron\bz{/}positron field, the photon field, and all other fields of the standard model of elementary particles) don't have such discrete substrate, hence they are continuous fields. All continuous quantum fields are elementary fields, and all elementary quantum fields are continuous fields. The notions \al continuous\ar  and \al elementary\ar   are interchangeable in case of quantum fields, and both will be used in this article. 

Strong experimental indications for a non vanishing zero\bz{-}point energy of phonon fields were available already in 1925 due to Mullican's evaluations of molecular vibrational spectra\!\cite{Mulliken:bospectr}. Few years later (in 1928) the zero\bz{-}point vibrations of rock\bz{-}salt were observed\!\cite{James:NaClzeropten}, and in the following decades the zero\bz{-}point energy of phonon fields has been experimentally confirmed beyond doubt\!\cite{Wilks:Helium}. 

Thus the assumption of a non vanishing zero\bz{-}point energy seems to be correct at least in case of phonon fields. If this assumption should be \emph{not} correct --- as I will argue in this article --- in case of elementary fields, this would mean that elementary fields differ basically from discrete fields, and consequently require a basically different treatment in quantum field theory. 

No consensus on the reality of a non vanishing zero\bz{-}point energy of elementary fields could yet been reached in the community of physicists. Exactly four \textsc{pro} arguments have been forwarded for the assumption of a diverging (respectively --- after regularization --- very huge) zero\bz{-}point energy of elementary quantum fields: 
\begin{ggitemize}
\ggitem{The first \textsc{pro} argument: As the non vanishing zero\bz{-}point energy of discrete fields has been experimentally confirmed, it is quite natural to expect that continuous fields should have the same property.}
\ggitem{The second \textsc{pro} argument: Casimir postulated a force, which is exerted (in the framework of his model) by the zero\bz{-}point oscillations of the electromagnetic field. This force has been experimentally confirmed.}
\ggitem{The third \textsc{pro} argument: The successful model of electro-weak interactions seems to indicate, that a phase change of the Higgs field has happened in the past. That phase change requires a non vanishing zero-point energy of the Higgs field.}
\ggitem{The fourth \textsc{pro} argument: The most widespread explanation for the cosmic inflation, which probably happened shortly after the big bang, assumes a phase change of an inflaton field. That phase change requires a non vanishing zero\bz{-}point energy of the inflaton field.}
\end{ggitemize}
And exactly three \textsc{contra} arguments have been forwarded against the assumption of a non vanishing zero\bz{-}point energy of elementary quantum fields: 
\begin{ggitemize}
\ggitem{The first \textsc{contra} argument: The indeterminacy relations enforce zero\bz{-}point oscillations in case of discrete fields, but not in case of elementary fields.}
\ggitem{The second \textsc{contra} argument: A non vanishing zero\bz{-}point energy of elementary fields should have a significant gravitational effect. This effect is not observed.}
\ggitem{The third \textsc{contra} argument: A non vanishing zero\bz{-}point energy of elementary fields is not compatible with special relativity theory.}
\end{ggitemize}
In the sequel, I will discuss and assess these seven arguments one by one. To make sure that we know what we are talking about, upfront briefly some essential facts about the ES\bz{-}tensor of quantum fields are recapitulated.

\section{The ES-tensor of quantum fields}\label{sec:estensor}
The abbreviation ES\bz{-}tensor will be used in this article for the energydensity\bz{-}stress\bz{-}tensor (sometimes imprecisely called energy\bz{-}momentum\bz{-}tensor) 
\begin{align}
(\mathcal{T}_{\mu\nu})=
\begin{pmatrix}
\mathcal{H}			& \ c\mathcal{P}_1 	& \ c\mathcal{P}_2	& \ c\mathcal{P}_3 \\
\mathcal{S}^{E}_1/c & \ \mathcal{S}^{P1}_1 & \ \mathcal{S}^{P2}_1 & \ \mathcal{S}^{P3}_1 \\
\mathcal{S}^{E}_2/c & \ \mathcal{S}^{P1}_2 & \ \mathcal{S}^{P2}_2 & \ \mathcal{S}^{P3}_2 \\
\mathcal{S}^{E}_3/c & \ \mathcal{S}^{P1}_3 & \ \mathcal{S}^{P2}_3 & \ \mathcal{S}^{P3}_3 \end{pmatrix} 
\end{align}
of (classical or quantum) fields. $\mathcal{H}$ is the field's energy density (\ggie  it's Hamiltonian), $\mathcal{P}_j$ is the $j$\bz{-}component of it's momentum density, $\mathcal{S}^{E}_k$ is the $k$\bz{-}component of the stream of energy density, and $\mathcal{S}^{Pj}_k$ is the $k$\bz{-}component of the stream of the $j$\bz{-}component of momentum density. 

I will argue that the ES\bz{-}tensor of elementary quantum fields differs from the ES\bz{-}tensor of discrete quantum fields and classical fields. According to conventional textbook wisdom, however, the ES\bz{-}tensor can be derived from the Lagrangian $\mathcal{L}$ due to the same formula in case of
\begin{subequations}\label{kdkjgdjnfdng}\begin{align}
&\hspace{-12mm}\text{classical fields, and arbitrary quantum fields:}\notag\\ 
\mathcal{T}_{\mu\nu}&=\frac{\partial\mathcal{L}}{\partial (\partial ^\mu\phi _\rho )}\,\partial _\nu\phi _\rho +\partial _\nu\phi ^* _\rho\,\frac{\partial\mathcal{L}}{\partial (\partial ^\mu\phi ^*_\rho )}-g_{\mu\nu}\mathcal{L}\label{kdkjgdjnfdnga} 
\end{align} 
$\phi (x)$ is the (classical or quantum) field. If $\phi ^*\! =\phi $, then the second term on the rhs is skipped. In the particular case of the Dirac field, $\phi ^*$ is replaced by $\phi ^\dagger \gamma ^{\; 0}$. $(g_{\mu\nu})$ is the metric tensor. Greek characters $\mu ,\nu ,\rho ,\dots $ are space\bz{-}time indices. They have to be summed over $0,1,2,3$ automatically, whenever they show up twice in a product. For space\bz{-}like indices latin characters $j,k,l,\dots $ will be used, which have to be summed over $1,2,3$ automatically whenever they show up twice in a product. 


In case of canonically quantized boson (fermion) fields $\phi (x)$, the volume integral over the ES\bz{-}tensor components \eqref{kdkjgdjnfdnga} becomes\,\footnote{\label{fn:textbook}See any textbook on quantum field theory, \ggeg  \cite[ch.\,2\,--\,4]{Peskin:QFT} or \cite[ch.\,14\,--\,17]{Gruendler:fieldtheory}.} 
\begin{align}
\inte _\Omega\!\dif ^3\! x\;\mathcal{T}_{\mu\nu}&=F_3\sum _{\boldsymbol{k},r}\frac{c\hbar k_\mu k_\nu }{\sqrt{k_jk^j+m^2c^2/\hbar ^2}}\,\cdot\notag\\ 
&\qquad\cdot\Big(\, ^r\! a^\dagger _{\boldsymbol{k}}\, ^r\! a_{\boldsymbol{k}}
+\, ^r\! b^\dagger _{\boldsymbol{k}}\, ^r\! b_{\boldsymbol{k}}
\,\pklammerm\,\underbrace{( ^r\! b_{\boldsymbol{k}}\, ^r\! b^\dagger _{\boldsymbol{k}}
\mklammerp\, ^r\! b^\dagger _{\boldsymbol{k}}\, ^r\! b_{\boldsymbol{k}})}_{1}
\Big)\, +\notag\\ 
&\qquad +(\text{nonlinear\,terms})-g_{\mu\nu}\Omega V_0\label{kdkjgdjnfdngb}\\ 
&\hspace{-8mm}\text{\textsc{if }} \phi ^*\!\neq\phi \text{ \textsc{then} }(F_3=1\ ,\ ^r\! b_{\boldsymbol{k}}\neq{}^r\! a_{\boldsymbol{k}})\notag\\ 
&\hspace{-8mm}\text{\textsc{else} }(F_3=1/2\ ,\ ^r\! b_{\boldsymbol{k}}={}^r\! a_{\boldsymbol{k}})\ .\notag 
\end{align}  
$\pklammerm $ is $+$ in case of boson fields, and $-$ in case of fermion fields. $\mklammerp $ is $-$ in case of boson fields, and $+$ in case of fermion fields. $m$ is the rest mass of a field quantum. $\sum _{\,\boldsymbol{k}}$ is the sum over all non\bz{-}redundant wavenumbers $\boldsymbol{k}$, which are compatible with the boundary conditions of the finite normalization volume $\Omega $. If an infinite normalization volume is chosen, then the sum over $\boldsymbol{k}$ is replaced by an integral. \al Non\bz{-}redundant\ar  means that in case of discrete fields the summation is only over the first Brillouin zone, \ggie  only over a finite number of wavenumbers. Continuous quantum fields, on the other hand, can have arbitrarily large wavenumbers. Hence the sum runs over infinitely many different wavenumbers in case of continuous fields, no matter whether a finite or an infinite normalization volume is chosen. 

For each wavenumber $\boldsymbol{k}$ and each polarization $r$ there is one particle oscillator with creation operator \raisebox{0mm}[0mm][0mm]{$\, ^r\! a^\dagger _{\boldsymbol{k}}$} and annihilation operator \raisebox{0mm}[0mm][0mm]{$\, ^r\! a_{\boldsymbol{k}}\, $}, and in case of complex fields $\phi ^*\!\neq\phi $ one anti\bz{-}particle oscillator with creation operator \raisebox{0mm}[0mm][0mm]{$\, ^r\! b^\dagger _{\boldsymbol{k}}$} and annihilation operator \raisebox{0mm}[0mm][0mm]{$\, ^r\! b_{\boldsymbol{k}}\, $}. If the field is real ($\phi ^*\!=\phi $), then \raisebox{0mm}[0mm][0mm]{$\, ^r\! b_{\boldsymbol{k}}={}^r\! a_{\boldsymbol{k}}\, $}. Examples for complex quantum fields are the fermion fields of the standard model of elementary particles. Examples for real quantum fields are the phonon field, the electromagnetic field, the Higgs field. 

The \al nonlinear terms\ar  are nonlinear in the particle\bz{-}number operators \raisebox{0mm}[0mm][0mm]{${}^r\! a^\dagger _{\boldsymbol{k}}\, ^r\! a_{\boldsymbol{k}}$} and \raisebox{0mm}[0mm][0mm]{${}^r\! b^\dagger _{\boldsymbol{k}}\, ^r\! b_{\boldsymbol{k}}$}\,. As in this article the focus is on the zero\bz{-}point values of ES\bz{-}tensors, these terms can be safely ignored, because they are negligible nearby the energy minimum of any oscillator. In case of all fields of the standard model of elementary particles with exception of the Higgs field, the nonlinear terms anyway are exactly zero. 

The constant energy\bz{-}density offset $V_0$ in the last term is different from zero only in case of fields with spontaneously broken symmetry. The Higgs field and the hypothetical inflaton field are important examples for such fields. 

The vacuum $|0\rangle $ is by definition that state, in which the expectation values of all particle\bz{-}number operators are zero, \ggie  the state in which no quantum at all is excited: 
\begin{align}
\langle 0|\,^r\! a^\dagger _{\boldsymbol{k}}\, ^r\! a_{\boldsymbol{k}}|0\rangle 
=\langle 0|\,^r\! b^\dagger _{\boldsymbol{k}}\, ^r\! b_{\boldsymbol{k}}|0\rangle =0\quad\forall\ \boldsymbol{k},r\notag 
\end{align} 
Due to the $\text{(anti)commutator}=\pklammerm 1$, and furthermore due to the $V_0$\bz{-}term in case of fields with spontaneously broken symmetry, the expectation value of \eqref{kdkjgdjnfdngb} never becomes zero, not even in the vacuum state. Instead the vacuum expectation values of the ES\bz{-}tensor components of a field $\phi $ are\pagebreak[1] 
\begin{align}
\langle 0|\,\mathcal{T}_{\mu\nu}|0\rangle &=\frac{F}{\Omega}\sum _{\boldsymbol{k}}\frac{c\hbar k_\mu k_\nu }{\sqrt{k_jk^j+m^2c^2/\hbar ^2}}-g_{\mu\nu}V_0\label{pogfkgkjmgfj}\\ 
F&=F_1\cdot F_2\cdot F_3\notag\\ 
F_1&=+1\text{ for bosons}\ ,\ F_1=-1\text{ for fermions}\notag\\ 
F_2&=\text{number of the field's polarization degrees of freedom}\notag\\  
F_3&=1\text{ \textsc{if} }\phi ^*\!\neq\phi\ ,\ F_3=1/2\text{ \textsc{else} .}\notag 
\end{align}\end{subequations} 
The diagonal elements of $\langle 0|\,\mathcal{T}_{\mu\nu}|0\rangle $ diverge in case of elementary fields, no matter whether the normalization volume is chosen finite or infinite, because $\sum _{\,\boldsymbol{k}}$ runs over infinitely many wavenumbers. 

In this article I will argue, that \eqref{kdkjgdjnfdng} is correct in case of discrete quantum fields (and, of course, in case of classical fields), but wrong in case of continuous quantum fields. This means that a split of \eqref{kdkjgdjnfdng} is suggested for the different cases of discrete and continuous quantum fields:
\begin{subequations}\label{kjsdjmnsnsdg}\begin{align}
&\hspace{-12mm}\text{classical fields, and discrete quantum fields:}\notag\\ 
\mathcal{T}_{\mu\nu}&=\frac{\partial\mathcal{L}}{\partial (\partial ^\mu\phi _\rho )}\,\partial _\nu\phi _\rho +\partial _\nu\phi ^\dagger _\rho\,\frac{\partial\mathcal{L}}{\partial (\partial ^\mu\phi ^\dagger _\rho )}-g_{\mu\nu}\mathcal{L}\label{kjsdjmnsnsdga}\\ 
&\hspace{-12mm}\text{elementary quantum fields:}\notag\\ 
\mathcal{T}_{\mu\nu}&=\frac{\partial\mathcal{L}}{\partial (\partial ^\mu\phi _\rho )}\,\partial _\nu\phi _\rho +\partial _\nu\phi ^\dagger _\rho\,\frac{\partial\mathcal{L}}{\partial (\partial ^\mu\phi ^\dagger _\rho )}-g_{\mu\nu}\mathcal{L}-Y\label{kjsdjmnsnsdgb}\\ 
Y&\equiv\,\text{the sum of all terms in \eqref{kjsdjmnsnsdga} which do not depend on}\notag\\  
&\hspace{5mm}\raisebox{.8mm}[0mm][3mm]{$\text{the particle\bz{-}number operators }^r\! a^\dagger _{\boldsymbol{k}}\, ^r\! a_{\boldsymbol{k}}\text{ or }\,^r\! b^\dagger _{\boldsymbol{k}}\, ^r\! b_{\boldsymbol{k}}$}\notag 
\end{align} 
With this split rule, the ES\bz{-}tensor of canonically quantized fields becomes 
\begin{align}
&\inte _\Omega\!\dif ^3\! x\;\mathcal{T}_{\mu\nu}\! =F_3\sum _{\boldsymbol{k},r}\frac{c\hbar k_\mu k_\nu }{\sqrt{k_jk^j+m^2c^2/\hbar ^2}}\,\cdot\notag\\ 
&\hspace{2.5mm}\cdot\Big(\, ^r\! a^\dagger _{\boldsymbol{k}}\, ^r\! a_{\boldsymbol{k}}
+\, ^r\! b^\dagger _{\boldsymbol{k}}\, ^r\! b_{\boldsymbol{k}}\,\pklammerm\,\underbrace{( ^r\! b_{\boldsymbol{k}}\, ^r\! b^\dagger _{\boldsymbol{k}}\mklammerp\, ^r\! b^\dagger _{\boldsymbol{k}}\, ^r\! b_{\boldsymbol{k}})}_{1}\,\Big)\, +\notag\\ 
&\hspace{2.5mm}+(\text{nonlinear\,terms})-g_{\mu\nu}\Omega V_0\quad \text{in case of discrete fields}\label{kjsdjmnsnsdgc}\displaybreak[1]\\ 
&\inte _\Omega\!\dif ^3\! x\;\mathcal{T}_{\mu\nu}\! =F_3\sum _{\boldsymbol{k},r}\frac{c\hbar k_\mu k_\nu }{\sqrt{k_jk^j+m^2c^2/\hbar ^2}}\Big(\, ^r\! a^\dagger _{\boldsymbol{k}}\, ^r\! a_{\boldsymbol{k}}
+\, ^r\! b^\dagger _{\boldsymbol{k}}\, ^r\! b_{\boldsymbol{k}}\Big)\, +\notag\\ 
&\hspace{2.5mm}+(\text{nonlinear\,terms})\quad\text{in case of continuous fields}\label{kjsdjmnsnsdgd}\\ 
&\hspace{2mm}\text{\textsc{if }} \phi ^*\!\neq\phi \text{ \textsc{then} }(F_3=1\ ,\ ^r\! b_{\boldsymbol{k}}\neq{}^r\! a_{\boldsymbol{k}})\notag\\ 
&\hspace{2mm}\text{\textsc{else} }(F_3=1/2\ ,\ ^r\! b_{\boldsymbol{k}}={}^r\! a_{\boldsymbol{k}})\notag 
\end{align}\end{subequations} 

While \eqref{kjsdjmnsnsdga} is the unchanged rule of classical field theory, the removal of all terms in \eqref{kjsdjmnsnsdgb} which do not depend on the particle\bz{-}number operators is an ad\bz{-}hoc postulate, \ggie  a law of nature. Like any law of nature, it was not derived but found by guessing, and can be justified by nothing else than the fact that it correctly reflects all experimental and observational experience. 

As an alternative to \eqref{kjsdjmnsnsdgb}, the measure of \al normal order\arp\cite[sec.\,4.3]{Peskin:QFT} is known since decades: 
\begin{align}
^r\! b_{\boldsymbol{k}}\, ^r\! b^\dagger _{\boldsymbol{k}}\mklammerp\, ^r\! b^\dagger _{\boldsymbol{k}}\, ^r\! b_{\boldsymbol{k}}\ \xrightarrow{\text{normal\,order}}\ \pklammerm{}^r\! b^\dagger _{\boldsymbol{k}}\, ^r\! b_{\boldsymbol{k}}\mklammerp\, ^r\! b^\dagger _{\boldsymbol{k}}\, ^r\! b_{\boldsymbol{k}}=0\notag 
\end{align} 
But normal order would not remove the term $g_{\mu\nu}\Omega V_0$\,. Furthermore normal order is merely a formal trick, while \eqref{kjsdjmnsnsdg} is corroborated by a plausible physical argument (the first \textsc{contra} argument), which will be presented in section\;\ref{sec:firstcontra}\,. 

The vacuum expectation value of the ES\bz{-}tensor $\eqref{kjsdjmnsnsdga}=\eqref{kjsdjmnsnsdgc}$ of discrete quantum fields is different from zero, and the vacuum expectation value of the ES\bz{-}tensor $\eqref{kjsdjmnsnsdgb}=\eqref{kjsdjmnsnsdgd}$ of continuous quantum fields is zero. In contrast, the vacuum expectation values of the ES\bz{-}tensors of both types of fields are different from zero, if $\eqref{kdkjgdjnfdnga}=\eqref{kdkjgdjnfdngb}$ should be correct for all types of quantum fields. Thus, if we want to find out whether \eqref{kdkjgdjnfdng} or \eqref{kjsdjmnsnsdg} is a correct description of quantum fields, we must answer the question: Is the vacuum actually filled by a non vanishing zero\bz{-}point energy of elementary fields, or not? 

\section{The first {\small PRO} argument} 
\emph{As the non vanishing zero-point energy of discrete fields has been experimentally confirmed, it is quite natural to expect that continuous fields should have the same property.}\vspace{.3\baselineskip}\\ 
This argument is obvious and easy to understand. No further explanation is needed. 

\section{The second {\small PRO} argument}\label{sec:caseff} 
\emph{Casimir postulated a force, which is exerted (in the framework of his model) by the zero-point oscillations of the electromagnetic field. This force has been experimentally confirmed.}\vspace{.3\baselineskip}\\ 
Casimir\!\cite{Casimir:caseffect} considered two perfectly conducting metal plates with area $A$, which are aligned parallel at distance $R$. If zero\bz{-}point oscillations of the electromagnetic field really exist, then some of these oscillations with wavelength $\lambda > R$ will be suppressed in the space between the plates. Hence the pressure of the zero\bz{-}point oscillations in\bz{-}between the plates will be slightly less than the pressure of the zero\bz{-}point oscillations in the outside space. Casimir computed that there should be a small net force 
\begin{align}
F_{\text{Casimir}}=-\frac{\pi ^2\hbar c\, A}{240\, R^4}\label{sdkdfbngbfc} 
\end{align}
pressing the plates towards each other. This force has been confirmed in many experiments. 
See \cite{Lambrecht:caseff2} for a review of the theory and experimental proofs of the Casimir effect. Many other phenomena can as well be computed in fair approximation due to the assumption of zero\bz{-}point oscillations of the electromagnetic field, for example the Lamb shift \cite{Martin:ccp}. 

On the other hand, all these phenomena can alternatively be computed by methods, which do not assume zero\bz{-}point oscillations of elementary fields: Casimir forces between macroscopic surfaces can be computed as van\bz{\:}der\bz{\:}Waals interactions\footnote{Van\bz{\:}der\bz{\:}Waals interactions are caused by fluctuating dipole moments of atoms, molecules, or solids\,\cite{London:molekkrae}. Atoms, molecules, and solids are \emph{not} elementary fields. Their fluctuating dipole moments are not related to zero\bz{-}point oscillations of the electromagnetic field, nor to zero\bz{-}point oscillations of any other elementary field.}, without invoking zero\bz{-}point oscillations of the electromagnetic field, as demonstrated by Lifshitz\,et.\,al. \cite{Lifshitz:vanderwaal,Dzyaloshinskii:vanderwaal}. And the Lamb shift can be computed\!\cite{Eides:hydrogatoms} by the standard perturbative methods of QED, again without invoking zero\bz{-}point oscillations of the electromagnetic field. 

Schwinger, DeRaad, and Milton\!\cite{Schwinger:casindie} demonstrated, that Casimir's and Lifshitz' computational methods lead to \emph{exactly} identical mathematical results for the Casimir force. Therefore experiments with increased precision will not tell us which method is a better description of nature. 

Thus we are faced with the fact that for all phenomena, which have been computed with reference to zero\bz{-}point oscillations of elementary fields, there exists an alternative computational method, which does not resort to those zero\bz{-}point oscillations. In a situation where two independent explanations exist for the same phenomena, one of them going without and one of them going with the assumption of zero\bz{-}point oscillations of elementary fields, clearly no stringent conclusion can be drawn regarding the reality of those zero\bz{-}point oscillations. Thus, from a logical point of view, we could stop the discussion at this point and simply state, that the experimental confirmation of the Casimir effect is not in conflict with the assumption, that zero\bz{-}point oscillations of elementary fields actually do not exist. Still a closer look on Casimir's method is appropriate, to assess the significance of this argument. 

Jaffe\!\cite{Jaffe:Casimir} named Casimir's reasoning \al heuristic\arp , thus characterizing this approach as a method which due to artful combination of assumptions, which are not sufficiently substantiated or even wrong, eventually arrives at correct results. The metal plates, which are attracted by the Casimir force, are in Casimir's method represented by boundaries with infinite conductivity. Thus at this point of his computations, Casimir implicitly assumed $\alpha\rightarrow\gginfty $ for the coupling constant $\alpha $ of the electromagnetic field, while in reality the conductivity of the metal plates is finite. But within the same computation he assumed that there is no exchange of virtual photons between the plates, \ggie  he implicitly assumed $\alpha\rightarrow 0$\,, while in reality virtual photons are coupling to fluctuating currents in the metal plates. Due to artful heuristic combination of these two wrong --- and extremely opposite --- implicit assumptions, Casimir eventually achieved the correct result. 

It's instructive to see how Casimir arrived at his method: In the fall of 1947, Casimir and Polder\cite{Casimir:polder} computed the retarded van\;der\;Waals force between two atoms without permanent dipole moments at large distance $R$. As a preparatory first step, they investigated a simpler setup, in which a single atom with polarizability $\beta $ is placed at a distance $R$ from a metal plane with infinite conductivity. Casimir and Polder computed the attractive force 
\begin{align}  
F=-\frac{3\hbar c\beta }{2\pi R^5}\label{sdkdfbngbfa}
\end{align}  
between the atom and it's mirror picture. In the next step, they found 
\begin{align}  
F=-\frac{161\hbar c\beta _1\beta _2}{4\pi R^8}\label{sdkdfbngbfb} 
\end{align}  
for the retarded van\;der\;Waals force between two atoms with polarizabilities $\beta _1$ and $\beta _2$. While the computation of \eqref{sdkdfbngbfb} was very complicated and tedious, Casimir and Polder were surprised by the simplicity, with which they had arrived at the result \eqref{sdkdfbngbfa}. Hence Casimir wondered whether the force could be computed with similar simplicity, if \emph{both} atoms were replaced by metal plates with infinite conductivity, \ggie  by boundaries. But while in case of \eqref{sdkdfbngbfa} the fluctuating dipole moment of the atom had supplied the electromagnetic field which is inducing the van\;der\;Waals interaction, where should the field come from in case of two boundaries? There are no fluctuating currents in boundaries. Casimir got the essential hint when he was chatting in those days with Bohr about his actual activities. \al Bohr mumbled something about zero\bz{-}point energy\arp , remembered Casimir many years later\cite{Casimir:bohr}. This tip was sufficient for Casimir, to find out that the electromagnetic field's zero\bz{-}point oscillations could replace the missing fluctuating currents, and to compute the force \eqref{sdkdfbngbfc}. 

Nowhere in the derivations of \eqref{sdkdfbngbfa} and \eqref{sdkdfbngbfb} had Casimir and Polder resorted to zero\bz{-}point oscillations of the electromagnetic field. And the derivation of \eqref{sdkdfbngbfc} is mathematically almost identical to the derivation of \eqref{sdkdfbngbfa}. Thus Casimir actually did nothing essentially different in the derivation of \eqref{sdkdfbngbfc}, but only replaced the fluctuating and\bz{/}or induced dipoles of the atoms by the assumed zero\bz{-}point oscillations of the electromagnetic field. 

(Virtual) photons interact with electrical charges, but zero\bz{-}point oscillations of the electromagnetic field don't. Hence the interaction between the two metal plates was lost when Casimir skipped the virtual photons from the picture (\ggie  implicitly assumed $\alpha\rightarrow 0$), and only kept the zero\bz{-}point oscillations. But the interaction was re\bz{-}gained when he in addition changed the metal plates to boundaries (\ggie  implicitly assumed $\alpha\rightarrow\gginfty $), because the spectrum of zero\bz{-}point oscillations is shaped by the geometry of the boundaries.  

While it is not obvious whether the field in\bz{-}between the plates are virtual photons or zero\bz{-}point oscillations, it's pretty clear and beyond doubt that the plates really are metal plates with finite conductivity, but not boundaries with infinite conductivity. Consequently the picture of virtual photons interacting with fluctuating currents in metal plates does correctly represent the actual situation, while the the picture with zero\bz{-}point oscillations inbetween boundaries is merely an artful substitution, which --- due to the intricate heuristic combination of $\alpha\rightarrow 0$ and $\alpha\rightarrow\gginfty $ --- leads to the same mathematical result. 

A precise analysis of Casimir's computational method has recently been published by Nikoli\'c\!\cite{Nikolic:Casimir,Nikolic:Casimir2}. His findings confirm, that the Casimir force actually is a van\bz{\:}der\bz{\:}Waals force, transmitted by virtual photons which couple to fluctuating currents in the metal plates, and not at all related to zero\bz{-}point oscillations of the electromagnetic field. 

\section{The first {\small CONTRA} argument}\label{sec:firstcontra} 
\emph{The indeterminacy relations enforce zero\bz{-}point oscillations in case of discrete fields, but not in case of elementary fields.}\vspace{.3\baselineskip}\\ 
The first \textsc{pro} argument says: Why should in case of continuous fields different laws of nature hold than in case of discrete fields? This \textsc{contra} argument is the direct answer. There is indeed a most important physical difference between discrete and continuous quantum fields: All discrete fields have material substrates, while all continuous quantum fields are substrate\bz{-}less fields. 

Consider for example the phonon field: It's material substrate is the atom grid of a solid or molecule. The state of the phonon field is at the same time the state of motion of the substrate particles. If $\langle 0|\,\mathcal{T}_{\mu\nu}^{\,\text{phonon}}\, |0\rangle $ would be zero for all $\mu $ and $\nu $, that would mean that the substrate particles would be at rest, thus having well\bz{-}defined positions and momenta at the same time, thus violating Heisenberg's indeterminacy relations\!\cite{Heisenberg:Unbestimmtheitsrel}. Therefore the basic tenets of quantum theory compellingly imply zero\bz{-}point oscillations in case of discrete quantum fields. 

Since Einstein introduced special relativity theory\!\cite{Einstein:SpezRelTheor}, we learned that no material substrate (the \al ether\arp ) can be assigned to the electromagnetic field, nor to any other elementary field. Elementary fields are substrate\bz{-}less fields. Hence there is no substrate which could come to rest, if no field quantum is excited. Consequently there is no conflict with quantum theory, if there are no zero\bz{-}point oscillations of substrate\bz{-}less fields. And we know that Nature uses to choose the most economical solutions for all her laws. Why should she all with a sudden waste without need infinite amounts of zero\bz{-}point energy to substrate\bz{-}less fields, while a much more economic solution, \ggie  the law \eqref{kjsdjmnsnsdg}, is easily at hand? 

All elementary fields are substrate\bz{-}less, and all substrate\bz{-}less fields are elementary. Thus we now have the three synonymous notions \al elementary\arp , \al continuous\arp , and \al substrate\bz{-}less\ar  for this type of quantum fields. The notion \al substrate\bz{-}less\ar  will be used frequently in the sequel, because it appropriately emphasizes that feature of elementary fields, which is the essential physical base for the law of nature \eqref{kjsdjmnsnsdgb}: If there exists no \al ether\arp , then the field has vanished to literally \emph{nothing}, if the expectation values \raisebox{0mm}[0mm][0mm]{$\langle\,{}^r\! a^\dagger _{\boldsymbol{k}}\, ^r\! a_{\boldsymbol{k}}\rangle$} and \raisebox{0mm}[0mm][0mm]{$\langle\,{}^r\! b^\dagger _{\boldsymbol{k}}\, ^r\! b_{\boldsymbol{k}}\rangle$} are zero for all $r$ and all $\boldsymbol{k}$. If no field quantum is excited, then there exists \emph{nothing} to which those contributions to the ES\bz{-}tensor could be assigned, which do not depend on the particle number operators \raisebox{0mm}[0mm][0mm]{${}^r\! a^\dagger _{\boldsymbol{k}}\, ^r\! a_{\boldsymbol{k}}$} or \raisebox{0mm}[0mm][0mm]{${}^r\! b^\dagger _{\boldsymbol{k}}\, ^r\! b_{\boldsymbol{k}}$}. Consequently such unphysical terms must be removed, as prescribed in \eqref{kjsdjmnsnsdgb}. 

\section{The second {\small CONTRA} argument}\label{sec:gravitation}
\emph{A non vanishing zero-point energy of elementary fields should have a significant gravitational effect. This effect is not observed.}\vspace{.3\baselineskip}\\ 
This contra argument was raised by Jordan and Pauli in an article\!\cite{Jordan:qedladfrei}, published in 1928, on the quantization of the electromagnetic field (my translation): \al It seems to us, that several considerations are indicating, that --- in contrast to the eigen\bz{-}oscillations in the crystal grid (where both theoretical and empirical reasons are indicating the existence of a zero\bz{-}point energy) --- no reality can be assigned to that `zero\bz{-}point energy'  $h\nu /2$ per degree of freedom in case of the eigen\bz{-}oscillations of the radiation. As one is dealing with regard to the latter with strictly harmonic oscillators, and as that `zero\bz{-}point radiation'  can neither be absorbed nor scattered nor reflected, it seems to elude, including it's energy or mass, any method of detection. Therefore it may be the simplest and most satisfactory conception, that in case of the electromagnetic field that zero\bz{-}point radiation does not exist at all.\ar\cite[page\,154]{Jordan:qedladfrei} 

With the words \al including it's energy or mass\arp , Jordan and Pauli are alluding to a conflict between the ES\bz{-}tensor \eqref{kdkjgdjnfdng} and general relativity theory (GRT). In the field equation\!\cite{Einstein:feldglann,Einstein:KosmolKonst} 
\begin{align}
R_{\mu\nu}(x)-\frac{R(x)}{2}\, g_{\mu\nu}(x)+\Lambda\, g_{\mu\nu}(x)=-\frac{8\pi G}{c^4}\,\mathcal{T}_{\mu\nu}(x) \label{lkdynjbgf}
\end{align}
of GRT, the Ricci\bz{-}tensor $(R_{\mu\nu})$ and it's contraction $R$ are representing the curvature of space\bz{-}time. $\Lambda $ is the cosmological constant, $G$ is the constant of gravitation, $(g_{\mu\nu})$ is the metric tensor, and $(\mathcal{T}_{\mu\nu})$ is the ES\bz{-}tensor of all fields existing at space\bz{-}time point $x$ with exception of the metric field $(g_{\mu\nu})$. 

Pauli famously\!\cite[page\,842]{Enz:Nullpkten} estimated (due to a cut\bz{-}off of short wavelengths at the classical electron radius) \al that the universe would not even reach to the moon\arp , if the electromagnetic field would really have a non vanishing zero\bz{-}point energy. By today, astronomical observations\!\cite{Bennet:WMAP9y,Planck_coll:cosmres2015} allow for a quite precise evaluation of the energy density in the universe: The universe can on large scales be well described by the \mbox{$\Lambda $CDM} model\!\cite{Bartelmann:cosmology} with dark energy parameter $\Omega _\Lambda =0.69$ and Hubble parameter $H=68\,\text{km}/(\text{s}\,\text{Mpc})$, resulting into the small vacuum energy density 
\begin{align} 
\mathcal{T}_{00}^{\,\text{vacuum}}=\frac{3H^2\Omega _\Lambda c^2}{8\pi G}=5.4\cdot 10^{-10} \,\mathrm{J/m^3}\ .\label{eq:energ2dichte}
\end{align}  
The observational data are furthermore indicating that the universe is in the vast empty regions far\bz{-}off mass concentrations an almost perfectly flat euclidean space. Therefore in this article the untypical isolated spots nearby mass concentrations with significant curvature of space\bz{-}time will be ignored. Instead only the typical areas of intergalactic vacuum will be considered, which are described by the metric 
\begin{align}
(g_{\mu\nu})\hspace{-.4em}\ggstackrel[1.3]{\stackrel{\scriptstyle\text{intergalactic}}{\text{vacuum}}}\hspace{-.4em}\text{diagonal}(+1,-a^2,-a^2,-a^2)\\ 
a(t)&=\text{cosmic scale factor}\ .\notag   
\end{align} 
Note that triple\bz{-}minus convention is chosen for the metric. With the usual choice $a(\text{today})\equiv 1$, and  \begin{align}
\frac{\dif a}{\dif t}\approx aH\approx 2\cdot 10^{-18}s^{-1} 
\end{align} 
being negligible on a timescale of, say, $10^{14}\text{\,s}\approx 3\text{\,million\,years}$, in excellent approximation Minkowski metric can be applied:  
\begin{align}
(g_{\mu\nu})\hspace{-.6em}\stackrel{\stackrel{\scriptstyle\text{intergalactic}}{\text{vacuum}}}{\approx}\hspace{-.6em}(\eta _{\mu\nu})=\text{diagonal}(+1,-1,-1,-1) 
\end{align} 

It's a plausible physical assumption that the vacuum is isotropic, and that consequently all off\bz{-}diagonal elements of $(\langle 0|\,\mathcal{T}_{\mu\nu}|0\rangle )$ vanish. Let $\text{P}\equiv\langle 0|\,\mathcal{T}_{11}|0\rangle =\langle 0|\,\mathcal{T}_{22}|0\rangle =\langle 0|\,\mathcal{T}_{33}|0\rangle $ be the isotropic vacuum pressure of an arbitrary elementary field. Now consider $(-\eta _{\mu\nu}\text{P})$. As $(\eta _{\mu\nu})$ is a Lo\-rentz\bz{-}covariant tensor and $-\text{P}$ is a Lo\-rentz\bz{-}invariant constant, $(\langle 0|\,\mathcal{T}_{\mu\nu}|0\rangle )$ impossibly could be a Lo\-rentz\bz{-}covariant tensor if it was identical with $(-\eta _{\mu\nu}\text{P})$ in all components with exception of $\langle 0|\,\mathcal{T}_{00}|0\rangle \neq -\eta _{00}\text{P}$. Thus combination of the requirement of Lorentz\bz{-}covariance and the assumption of isotropy of the vacuum implies 
\begin{align}
\langle 0|\,\mathcal{T}_{00}|0\rangle =-\langle 0|\,\mathcal{T}_{11}|0\rangle =-\langle 0|\,\mathcal{T}_{22}|0\rangle =-\langle 0|\,\mathcal{T}_{33}|0\rangle \ ,\label{iagnsaygrb} 
\end{align} 
as pointed out by Zeldovich\!\cite{Zeldovich:ccthelpart}. As $\langle 0|\,\mathcal{T}_{\mu\nu}\, |0\rangle /F=\eqref{pogfkgkjmgfj}/F$ diverges towards $+\gginfty $, but not towards $-\gginfty $, in all diagonal elements, this condition is not at all trivial. Zeldovich\!\cite{Zeldovich:ccthelpart} remarked that cut-off regularization, \ggie  replacing in case of elementary fields the summation limit $|\boldsymbol{k}|=\gginfty  $ in \eqref{pogfkgkjmgfj} by $|\boldsymbol{k}|_{\text{max}}=B<\gginfty $ and then considering $\lim _{B\rightarrow\ggginfty}$ is not appropriate, because this method does not change the signs of the spatial or time\bz{-}like terms, and consequently can not meet the condition \eqref{iagnsaygrb}. But he assumed that a relativistically covariant method of regularization would make \eqref{pogfkgkjmgfj} compatible with \eqref{iagnsaygrb}. 

In appendix\;\hyperlink{ta:appa}{A}\hspace{.3em} Zeldovich's assumption is checked and confirmed. There the results of covariant regularization 
\begin{subequations}\label{jmndnfgns}\begin{align} 
\langle 0|\,\mathcal{T}_{\mu\nu}\, |0\rangle\ggstackrel[.15]{\eqref{ksdjmghnfdx}}-\eta _{\mu\nu}\,\frac{F\hbar cK^{4}}{16\pi ^{2}}\lim _{w\rightarrow\ggginfty }\ln\!\Big(\frac{w}{K}\Big)\label{jmndnfgnsa}\\ 
\ggstackrel[-.1]{\eqref{msngnsdfnhgx}}-\eta _{\mu\nu}\,\frac{F\hbar c}{16\pi ^2}\lim _{\kappa\rightarrow\ggginfty }\kappa ^4\,\ln\!\Big(\frac{\kappa}{K}\Big)\label{jmndnfgnsb}\\ 
K&\equiv mc/\hbar >0\notag 
\end{align}\end{subequations} 
are derived, which meet the condition \eqref{iagnsaygrb}. \eqref{jmndnfgnsa} is the result of dimensional regularization, first published by Akhmedov\!\cite{Akhmedov:vacener}. \eqref{jmndnfgnsb} is the result of Pauli\bz{-}Villars regularization. It is explicated in appendix\;\hyperlink{ta:appa}{A}, why both regularization results are strictly infinite, even in case $m\rightarrow 0\, $. 

With some plausible assumptions, which are explicated in appendix\;\hyperlink{ta:appb}{B}, a finite range for the zero\bz{-}point energy density of an elementary field can be derived from \eqref{jmndnfgnsb}: 
\begin{align}
\langle 0|\,\mathcal{T}_{00}\, |0\rangle\stackrel{\eqref{mksmgmnsdfg}}{\approx}-F\cdot\Big(\, 5\cdot 10^{\, 48}\dots\, 4\cdot 10^{\, 111}\,\Big)\,\frac{\text{J}}{\text{m}^{3}}\label{kmsmjgnsdffg}
\end{align} 
Comparison of \eqref{kmsmjgnsdffg} with the vacuum energy density as concluded from astronomical observations, results into the disturbing ratio 
\begin{align}
\frac{\text{theory}}{\text{observation}}=\frac{\eqref{kmsmjgnsdffg}}{\eqref{eq:energ2dichte}}\approx -F\cdot\Big(\, 10^{\, 58}\dots\, 10^{\, 121}\,\Big)\ .\label{oirfkhdjhs} 
\end{align} 
The discrepancy can easily be absorbed by the cosmological constant. Simply assume 
\begin{align}
\Lambda\stackrel{\eqref{eq:energ2dichte},\eqref{lkdynjbgf}}{=}\frac{8\pi G}{c^4}\Big( 5.4\cdot 10^{-10}\, \mbox{Jm}^{-3}-\sum _z\langle 0|\,\mathcal{T}_{00}^{\, (z)}\, |0\rangle\Big) \label{syfhgjnkgfd} 
\end{align} 
with $\sum _{\, z}$ being the sum over all elementary fields (not including the metric field) existing in the universe, and everything is fine. We just need to believe that $\Lambda $ and \raisebox{0mm}[0mm][0mm]{$\sum _{\, z}\langle 0|\,\mathcal{T}_{00}^{\, (z)}\, |0\rangle $} really mutually compensate (by chance right now, in the present epoch of cosmic evolution) with the breathtaking accuracy of $58\dots 121$ decimal digits. That's of course hard to believe as long as nobody comes up with a physical explanation. 

It is much more plausible to assume that \eqref{kjsdjmnsnsdg} is a correct law of nature, and that consequently $\langle 0|\,\mathcal{T}_{00}\, |0\rangle =\eqref{kmsmjgnsdffg}$ actually is not ${\approx}-F\cdot (\, 5\cdot 10^{\, 48}\dots\, 4\cdot 10^{\, 111}\,)\,\text{J}/\text{m}^{3}$, but simply zero. This does of course not answer the question, why the universe is filled with the small energy density $\mathcal{T}_{00}^{\,\text{vacuum}}\,\textstackrel{\eqref{eq:energ2dichte}}\,5.4\cdot 10^{-10} \,\mathrm{J/m^3}$. But for this question we have anyway no answer, no matter whether we assume \eqref{kdkjgdjnfdng} or \eqref{kjsdjmnsnsdg} to be the correct law of nature. With \eqref{kjsdjmnsnsdg} we have at least the \emph{gigantic} advantage that the \emph{gigantic} mismatch \eqref{oirfkhdjhs} is removed.

At this point a remark is due on an influential article, which Weinberg\!\cite{Weinb:KosmConst} published in 1989 on the mismatch \eqref{oirfkhdjhs}. Weinberg believed in the reality of the energy density \eqref{kmsmjgnsdffg}, because he considered the Casimir force a valid proof for the reality of the electromagnetic field's zero\bz{-}point energy (the second \textsc{pro} argument), and because he believed in the historic fact of an electro\bz{-}weak phase transition (the third \textsc{pro} argument). Hence he was in need to find a plausible explanation for the gigantic mismatch \eqref{oirfkhdjhs}. He compiled many convincing arguments, why \eqref{syfhgjnkgfd}, though mathematically perfectly correct, is not at all a physically sound remedy, and dubbed the issue \al the cosmological constant problem\arp . 
In my point of view, there is no need for a solution of the cosmological constant problem, because that problem --- \ggie  the huge energy density \eqref{kmsmjgnsdffg} --- actually doesn't exist. 

\section{The third {\small PRO} argument}\label{sec:higgs} 
\emph{The successful model of electro-weak interactions seems to indicate, that a phase change of the Higgs field has happened in the past. That phase change requires a non vanishing zero-point energy of the Higgs field.}\vspace{.3\baselineskip}\\ 
The Glashow\bz{-}Salam\bz{-}Weinberg model\footnote{See for example \cite[chap.\,29]{Gruendler:fieldtheory} for an elementary introduction to the GSW\bz{-}model of electroweak interactions and the Higgs mechanism.} of electroweak interactions assumes the existence of a complex weak isospin\bz{-}doublet 
\begin{align} 
&\phi (x)\! =\!\frac{1}{\sqrt{2}}\!\begin{pmatrix}{\phi}_{3}(x)+i{\phi}_{4}(x)\\ {\phi}_{1}(x)+i{\phi}_{2}(x)\end{pmatrix}
\! =\!\frac{1}{\sqrt{2}}\!\begin{pmatrix}{\phi}_{3}(x)+i{\phi}_{4}(x)\\ \sqrt{2}\, f+{\chi}(x)+i{\phi}_{2}(x)\end{pmatrix}\label{mdngnsnbg}\\ 
&{\phi}_{1}(x),\,{\phi} _{2}(x),\,{\phi}_{3}(x),\,{\phi}_{4}(x),\,{\chi}(x)\in\mathbbm{R}\hspace{2mm},\hspace{2mm}0<f=\text{constant}\in\mathbbm{R}\notag 
\end{align} 
with Lagrangian 
\begin{subequations}\label{jsngnsdgbn}\begin{align}
\mathcal{L}&=c^2\hbar ^2(\Dif _\mu\phi ^\dagger )\Dif ^\mu\phi +\frac{m^2c^4}{2}\,\phi ^\dagger\phi -\frac{m^2c^4}{4f^2}\, (\phi ^\dagger\phi )^2\ .\label{kasbngjsdgfdx}
\end{align} 
The field's potential energy density, displayed as a solid curve in fig.\,\ref{fig:iosnjhgnhs}\,, is minimal at $|\phi |=f$. The covariant differential operators $\Dif _\mu $ bring the three weak gauge bosons $W^+$, $W^-$, $Z^0$ and the weak coupling constants $g_1$ and $g_2$ into play. 
\begin{figure}[!t]\centering\begin{overpic}[trim=0 -1mm 0 0]{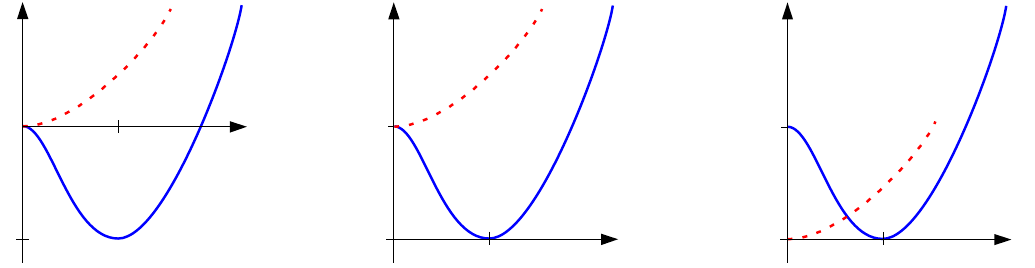}
\put(-8,25){\makebox[10mm][r]{$V$}}
\put(-2.5,19){\scalebox{1.3}{\ding{172}}}
\put(-8,13.5){\makebox[10mm][r]{$0$}}
\put(-8.3,2){\makebox[10mm][r]{$-V_0$}}
\put(23,11){$|\phi |$}
\put(10.5,11){$f$}
\put(5,21.7){\textcolor{red}{$\scriptstyle T>T_c$}}
\put(17,6.6){\textcolor{blue}{$\scriptstyle T<T_c$}}
\put(35.6,0){
\put(-8,25){\makebox[10mm][r]{$V$}}
\put(-2.5,19){\scalebox{1.3}{\ding{173}}}
\put(-8,13.7){\makebox[10mm][r]{$V_0$}}
\put(-8.3,2.4){\makebox[10mm][r]{$0$}}
\put(23,0){$|\phi |$}
\put(10.5,0){$f$}
\put(5,21.7){\textcolor{red}{$\scriptstyle T>T_c$}}
\put(17.2,6.6){\textcolor{blue}{$\scriptstyle T<T_c$}}
}
\put(73.6,0){
\put(-8,25){\makebox[10mm][r]{$V$}}
\put(-2.5,19){\scalebox{1.3}{\ding{174}}}
\put(-8,13.7){\makebox[10mm][r]{$V_0$}}
\put(-8.3,2.4){\makebox[10mm][r]{$0$}}
\put(23,0){$|\phi |$}
\put(10.5,0){$f$}
\put(8.5,14.5){\textcolor{red}{$\scriptstyle T>T_c$}}
\put(17.2,6.6){\textcolor{blue}{$\scriptstyle T<T_c$}}
}
\end{overpic}\caption{Solid line: Potential energy density of the Lagrangian \eqref{kasbngjsdgfdx}. Dashed line: Potential energy density of the Lagrangian \eqref{osksgnsfghs}. A constant offset of vacuum energy density $\neq 0$ exists \ding{172} after the phase transition, \ding{173} before the phase transition, \ding{174} never.}\label{fig:iosnjhgnhs}
\end{figure}
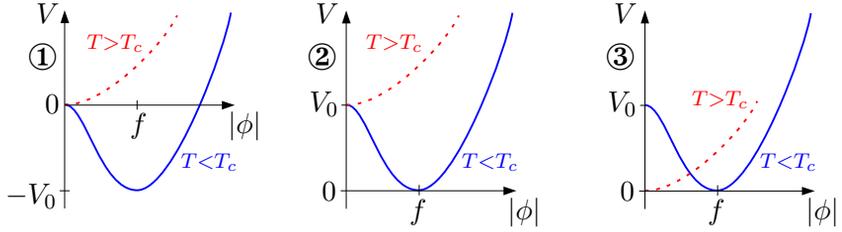

The Lagrangian \eqref{kasbngjsdgfdx} is invariant under arbitrary rotations in the four\bz{-}dimensional space spanned by $\phi _1$\bz{\,,\,}$\phi _2$\bz{\,,\,}$\phi _3$\bz{\,,\,}$\phi _4$\,, but the field \eqref{mdngnsnbg} is not. With no reason (\al spontaneously\arp ) one direction in this four\bz{-}dimensional space became distinguished due to $f$\,. Without loss of generality, in \eqref{mdngnsnbg} $\phi _1$\bz{\,,\,}$\phi _2$\bz{\,,\,}$\phi _3$\bz{\,,\,}$\phi _4$ have been rotated such, that $f$ coincides with the $\phi _1$\bz{-}direction. 

The three Goldstone bosons $\phi _2$\bz{\,,\,}$\phi _3$\bz{\,,\,}$\phi _4$ are not physical. This can be made explicitly visible due to an appropriate gauge transformation. The Lagrangian \eqref{kasbngjsdgfdx} thereby is transformed to 
\begin{align}
&\mathcal{L}\! =\!\frac{c^2\hbar ^2}{2}\, (\mbox{d}_\mu{\chi})^\dagger \mbox{d}^\mu{\chi}\!  
-\! \frac{m^2c^4}{2}\,\chi ^2\! -\!\frac{m^2c^4}{f\sqrt{8}}\,\chi ^3\! -\!\frac{m^2c^4}{16f^2}\,\chi ^4\! +\!\underbrace{\frac{m^2c^4f^2}{4}}_{V_0}+\notag\\ 
&+\frac{g_2^2f^2}{4c^2}\,c^4\,(W^{+}_{\mu}W^{+\mu}+W^{-}_{\mu}W^{-\mu})+\frac{(g_1^2+g_2^2)f^2}{4c^2}\,c^4\,Z^{0}_{\mu}Z^{0\mu}\ ,\label{jsngnsdgbnc}
\end{align}\end{subequations} 
and the field \eqref{mdngnsnbg} is transformed to 
\begin{align} 
&\phi (x)=\frac{1}{\sqrt{2}}\begin{pmatrix}\, 0\, +i\, 0\,\\ \phi _1(x)+\, i\, 0\,\end{pmatrix}
=\frac{1}{\sqrt{2}}\begin{pmatrix}0\\ \sqrt{2}\, f+\chi (x)\end{pmatrix}\ .\label{kjmdmndfgnsd}
\end{align} 
The Goldstone bosons $\phi _2$\bz{\,,\,}$\phi _3$\bz{\,,\,}$\phi _4$ have disappeared, the remaining massive field $\chi\equiv\phi _1-\sqrt{2}\, f$ oscillates around the minimum of the potential energy at $\phi _1=\sqrt{2}f$, and the weak gauge bosons $W^+$, $W^-$, $Z^0$ became massive (note the parameter $f\neq 0$ in their mass terms). Without coupling to the Higgs field, the three weak gauge bosons would be mass\bz{-}less, and would consequently have only two transversal polarization degrees of freedom like the photon and the gluons. As they became massive now, they have each the additional degree of freedom of longitudinal polarization. Thus here the three degrees of freedom, which vanished with the three Goldstone bosons, are showing up again. 

The Lagrangian \eqref{jsngnsdgbnc} not only explains why the gauge bosons are massive, it also predicts the ratio 
\begin{align}
\frac{\text{mass}(W^\pm )}{\text{mass}(Z^0)}\approx\frac{g_2}{\sqrt{g_1^2+g_2^2}}\ .
\end{align}
This prediction (which is not exact, because the mass terms in \eqref{jsngnsdgbnc} are valid only on tree\bz{-}level, and must be amended by higher order perturbative corrections) was verified, when in 1983 first time $W^{\pm}$ bosons \cite{UA1:Wdetection,UA2:Wdetection} with mass $80\,\text{Gev}/c^2$, and $Z^0$ bosons \cite{UA1:Zdetection} with mass $91\,\text{Gev}/c^2$ were produced and observed in collider experiments. 

Both $\chi (x)$ and $\phi (x)$ are commonly called Higgs field. This boson field with mass $m=125\,\text{Gev}/c^2$ has been observed first time in 2012 \cite{Atlas:Higgs2012,CMS:Higgs2012}. Thus an impressive amount of experimental evidence is indicating, that the Lagrangian \eqref{jsngnsdgbn} is a correct description of nature. 

Due to measurements of myon decay, the coupling constant $g_2$ could be disentangled from the $W^\pm $ mass term. Thereby the parameter 
\begin{align}
f=246\,\text{GeV}(\hbar c)^{-3/2}
\end{align}
could be computed. Inserting this value into the constant term in \eqref{jsngnsdgbnc}, we get 
\begin{align}
V_0=\frac{m^2c^4f^2}{4}=\frac{(125\,\text{GeV})^2(246\,\text{GeV})^2}{4(\hbar c)^{3}}\approx 2\cdot 10^{\, 34}\,\frac{\text{J}}{\text{m}^3}\label{ksmndgnsdnsgf}
\end{align}
While this energy density is many orders of magnitude less than \eqref{kmsmjgnsdffg}, the ratio 
\begin{align}
\frac{\text{theory}}{\text{observation}}=\frac{\eqref{ksmndgnsdnsgf}}{\eqref{eq:energ2dichte}}\approx 3.7\cdot 10^{\, 43}\label{ksmjsgnsfg} 
\end{align} 
is still terrifying. The problem can easily be removed: We merely need to assume that not the sketch \ding{172} but the sketches \ding{173} and \ding{174} in fig.\;\ref{fig:iosnjhgnhs} correctly describe the actual situation. But this is just what the proposed law \eqref{kjsdjmnsnsdg} does. It removes the term $V_0$ from the ES\bz{-}tensor, because this term does not depend on any particle\bz{-}number operator. Thus at first sight this seems not to be a \textsc{pro} argument, but a \textsc{contra} argument. Still the issue is listed as a \textsc{pro} argument, for the following reason: 

The six inventors \cite{Englert:SymBrech,Guralnik:higgs,Higgs:symbrecha,Higgs:symbrechb} of the Lagrangian \eqref{jsngnsdgbn} would have hardly been able to establish such a complicated construction, if they had not been guided by the vision of an electro\bz{-}weak phase transition. The diagram $V(\phi )$ in fig.\,\ref{fig:iosnjhgnhs} resembles the diagram $V(M)$ of the free energy density of a ferromagnet with macroscopic magnetization $M$. At high temperature the macroscopic magnetization is zero. But when the temperature drops below the Curie temperature, the ferromagnet can lower it's free energy due to the formation of a finite macroscopic magnetization. In analogy we may imagine that the Higgs field $\phi (x)$ once was in a Klein\bz{-}Gordan phase with Lagrangian 
\begin{align} 
\mathcal{L}=c^2\hbar ^2(\mbox{d}_\mu \phi ^\dagger )\mbox{d}^\mu \phi -m^2c^4\phi ^\dagger\phi\ ,\label{osksgnsfghs}  
\end{align} 
whose potential energy density is displayed as the dashed curve in fig.\;\ref{fig:iosnjhgnhs}\,, and then changed due to an \al electroweak phase transition\ar  to the phase described by the Lagrangian \eqref{jsngnsdgbn}, when the temperature of the universe dropped below the critical value $T_c\approx 160\,\text{GeV}/k_B\approx 1.9\cdot 10^{\, 15}\text{K}$\cite{Onofrio:phasetrans} at about $10^{\, -12}\text{s}$ after the big bang. 

If \eqref{kjsdjmnsnsdg} is the correct law of nature, then such phase change of elementary fields is impossible, however, because there is no energy difference which could drive the phase change: If the Higgs field is in the Klein\bz{-}Gordan phase \eqref{osksgnsfghs}, and assumes the state $|0\rangle $ of lowest energy with $\phi =0$, then according to \eqref{kjsdjmnsnsdg} it's energy is zero. If the field is in the phase \eqref{jsngnsdgbn}, and assumes the state $|0\rangle $ of lowest energy with $\phi =\big(\begin{smallmatrix}0\\ f\end{smallmatrix}\big) $, then according to \eqref{kjsdjmnsnsdg} it's energy as well is zero, see sketch \ding{174} in fig.\;\ref{fig:iosnjhgnhs}\,. Thus the Higgs field can not lower it's energy by a phase change from the Klein\bz{-}Gordan phase into the phase \eqref{jsngnsdgbn}. If \eqref{kjsdjmnsnsdg} is correct, then the concept of phase changes is applicable only to fields with material substrates, but not to substrate\bz{-}less fields. 

Therefore this argument becomes a \textsc{pro} argument due to this line of reasoning: The Higgs Lagrangian \eqref{jsngnsdgbn} has been conclusively confirmed. This Lagrangian can be derived from the assumption of an electro\bz{-}weak phase change, hence this phase change must really have happened. But \eqref{kjsdjmnsnsdg} is not compatible with such phase change. Thus \eqref{kjsdjmnsnsdg} is disproved by the success of the Lagrangian \eqref{jsngnsdgbn}. 

This reasoning is wrong, however. The confirmation of the Lagrangian \eqref{jsngnsdgbn} does \emph{not} prove that some electroweak phase transition ever has happened. To understand the argument, read Maxwell's eloquent proof\!\cite{Maxwell:ether} of the reality of the luminiferous ether. Maxwell conceived his theory of the electromagnetic field based on a clear vision of some material substrate, which was to transport electromagnetic waves like a string is transporting mechanical wavy motions. He misunderstood the experimental confirmations of his theory of electrodynamics as evidence for the existence of the ether, and computed the amazing values of that ubiquitous material's density and modulus of elasticity. But the assumption of the ether resulted into unsurmountable problems, and eventually that concept was abandoned. Still Maxwell's equations are a valid and correct description of nature. 

Now compare the Higgs field and the macroscopic magnetic field $M$ of a ferromagnet. The constituent particles of the ferromagnetic solid, which have permanent magnetic dipole moments, are the material substrate of the magnetic field $M$. Sketch \ding{173} in fig.\;\ref{fig:iosnjhgnhs} describes the ferromagnet: At high temperature, the rotational energy of the constituent particles is high, hence no macroscopic magnetic field is formed. Below the Curie temperature, the rotations are frozen to mere vibrations around a spontaneously chosen preferred axis, and the energy density $V_0$, which had been stored in the rotations of the constituent particles, is dissipated to the environment. 

The Higgs field, in contrast, has no material substrate in which the energy density $V_0$ could be stored before or after a phase transition, and therefore the analogy with the magnetic field must not be taken literally. We must not repeat Maxwell's error and misinterpret the experimental confirmations of the Glashow\bz{-}Salam\bz{-}Weinberg model of electroweak interactions and the direct observation of the Higgs boson as confirmations of some Higgs\bz{-}ether, which would correspond to the ferromagnetic solid. These experimental confirmations prove not more nor less than the existence of a field, which is correctly described by \eqref{jsngnsdgbn}. 

Like the vision of the luminiferous ether helped Maxwell to construct his equations of electrodynamics, the vision of a phase change helped the inventors of the Higgs field to construct the Lagrangian \eqref{jsngnsdgbn}. Still neither the supposed existence of a luminiferous ether nor the supposed historical fact of an electroweak phase change are of any relevance for the validity of Maxwell's equations or the Higgs\bz{-}Lagrangian \eqref{jsngnsdgbn}. The only question that matters is, whether or not the consequences drawn from these equations are confirmed by all experimental observations. These equations are laws of nature, which stand at the very begin of the respective theories. They are not in need of derivation, nor in need of colorful justifications. More than a century after Einstein abandoned the ether, we should have learned to trust in the abstract mathematics of well\bz{-}proven laws of nature, and not insist on pictorial explanations.  

\section{The fourth {\small PRO} argument}\label{sec:inflation} 
\emph{The most widespread explanation for the cosmic inflation, which probably happened shortly after the big bang, assumes a phase change of an inflaton field. That phase change requires a non vanishing zero\bz{-}point energy of the inflaton field.}\vspace{.3\baselineskip}\\ 
Why is the visible universe as homogeneous, as isotropic, and as flat as we observe it by today? And why don't we observe magnetic monopoles? An elegant answer to all these questions is provided by the assumption, that the universe expanded during the short time interval from about $10^{\, -35}\text{s}$ to about $10^{\, -32}\text{s}$ after the big bang by a factor of minimum $10^{\, 26}$. This enormous expansion, called cosmic inflation, can furthermore explain the mass concentration in galaxies and galaxy clusters as a result of quantum fluctuations in the early universe, which were stretched during the inflation. Thus it is quite likely that a cosmic inflation really has happened shortly after the big bang. 

The most widespread models of inflation are assuming that the expansion was caused by the phase change of an inflaton field\cite{Linde:RevInflation}. As explained in the previous section, such models are not compatible with the proposed law of nature \eqref{kjsdjmnsnsdg}. According to this law, there are no phase changes of elementary quantum fields, because the vacuum energy of any elementary field is zero. Hence there is no energy difference, which could drive a phase change. Consequently \eqref{kjsdjmnsnsdg} is wrong, and instead \eqref{kdkjgdjnfdng} is valid for \emph{all} types of fields, including elementary (\ggie  substrate\bz{-}less) quantum fields, if the cosmic inflation really was caused by the phase change of an inflaton field. 

This model of inflation, however, is not without problems. As the universe is almost flat by today, the inflaton field must have had the huge energy offset $V_0$ before the phase change, \ggie  sketch \ding{173} in fig.\,\ref{fig:iosnjhgnhs} applies. Nobody could yet propose a plausible reason for such huge energy offset before the phase change. Furthermore the application of classical general relativity theory (with no quantization of space\bz{-}time) may be basically misleading so close to the \al big bang\ar  space\bz{-}time singularity. Computations in the framework of loop quantum gravity are indicating fundamental modifications of the inflation scenario\cite{Ashtekar:inflation}. In any case all models of inflation are (inevitably) quite speculative, because no doublecheck of the physics at such high energy density is possible in the laboratory. There are strong arguments for the historic fact of a cosmic inflation, but the cause for this event may very well be something completely different from the phase change of an elementary field. 

Furthermore a very basic weakness in the concept of phase changes of elementary fields can be hardly ignored: Something can change it's phase if and only if it has internal degrees of freedom, in which energy can be stored, and from which energy can be released in course of the phase change (the rotational degrees of freedom of the constituent particles of the ferromagnet are a good example). Being substrate\bz{-}less fields, elementary fields don't have internal degrees of freedom (that's why they are called elementary). Where, then, should they store the energy, which would be needed to drive a phase change? The thoughtless transfer of the concept of phase changes from material systems (\ggie  the substrates of discrete fields) to elementary fields may very well be an over\bz{-}simplification, and basically misleading. 

Considering this fundamental objection, and furthermore our quite incomplete knowledge of the physics prevailing close to the big bang, the inflaton\bz{-}phase\bz{-}change argument must be rated \al weak\arp . 

\section{The third {\small CONTRA} argument}\label{sec:signsincomp} 
\emph{A non vanishing zero-point energy of elementary fields is not compatible with special relativity theory.}\vspace{.3\baselineskip}\\ 
As explicated in section\;\ref{sec:gravitation}\,, the combination of the requirement of Lorentz\bz{-}covariance and the assumption of isotropy of the vacuum implies for the ES\bz{-}tensors of elementary (\ggie  continuous) fields 
\begin{subequations}\label{kjsdngnsdsg}\begin{align}
&\langle 0|\,\mathcal{T}_{00}^{\text{\,continuous}}|0\rangle\stackrel{\eqref{iagnsaygrb}}{=}-\langle 0|\,\mathcal{T}_{11}^{\text{\,continuous}}|0\rangle\, =\notag\\ 
&\hspace{2em}=-\langle 0|\,\mathcal{T}_{22}^{\text{\,continuous}}|0\rangle =-\langle 0|\,\mathcal{T}_{33}^{\text{\,continuous}}|0\rangle \ .\label{kjsdngnsdsga} 
\end{align} 
Note that the constant energy\bz{-}density offset $V_0$ showing up in the Higgs Lagrangian \eqref{jsngnsdgbnc} (and in the Lagrangian of the hypothetical inflaton field) is compatible with \eqref{kjsdngnsdsga}:
\begin{align} 
\langle 0|\,\mathcal{T}_{\mu\nu}^{\text{\,Higgs}}\, |0\rangle\stackrel{\eqref{pogfkgkjmgfj},\eqref{jmndnfgnsb}}{=}-\eta _{\mu\nu}\,\frac{F\hbar c}{16\Omega\pi ^2}\lim _{\kappa\rightarrow\ggginfty }\kappa ^4\,\ln\!\Big(\frac{\kappa}{K_0}\Big) -\eta _{\mu\nu}V_0\notag 
\end{align} 

On the other hand, the signs of all diagonal elements of the ES\bz{-}tensor are identical for discrete fields (unless the ES\bz{-}tensor should be dominated by the term $-\eta _{\mu\nu}V_0$, which is certainly not true in case of very many really existing discrete fields): 
\begin{align} 
&\text{sign}\Big(\langle 0|\,\mathcal{T}_{00}^{\text{\,discrete}}\, |0\rangle\Big)\stackrel{\eqref{pogfkgkjmgfj}}{=}\text{sign}\Big(\langle 0|\,\mathcal{T}_{11}^{\text{\,discrete}}\, |0\rangle\Big) =\notag\\
&=\text{sign}\Big(\langle 0|\,\mathcal{T}_{22}^{\text{\,discrete}}\, |0\rangle\Big) =\text{sign}\Big(\langle 0|\,\mathcal{T}_{33}^{\text{\,discrete}}\, |0\rangle\Big)\label{kjsdngnsdsgb} 
\end{align}\end{subequations} 

Now consider for example the ES\bz{-}tensor of the Higgs field (\ggie  a continuous field), and the ES\bz{-}tensor of the phonon field of a solid consisting of $N$ atoms (\ggie  a discrete field). Both are real scalar boson fields, hence for both the parameter $F\textstackrel{\eqref{pogfkgkjmgfj}}(+1)\cdot (1)\cdot (1/2)$ applies. Despite this similarity, \eqref{kjsdngnsdsga} is valid for the Higgs field, while \eqref{kjsdngnsdsgb} is valid for the phonon field, no matter how huge the number $N$ may be. But a change of signs somewhere in\bz{-}between \al arbitrary huge\ar  and \al infinite\ar  can not be reasonably justified. The difference of signs between \eqref{kjsdngnsdsga} and \eqref{kjsdngnsdsgb} would be an unacceptable inconsistency, unless the zero\bz{-}point expectation value of the ES\bz{-}tensor is zero in case of elementary fields and\bz{/}or in case of discrete fields. The non vanishing zero\bz{-}point energy of phonon fields has been experimentally observed and confirmed beyond doubt\cite{Wilks:Helium}. Hence we must have 
\begin{align}
\langle 0|\,\mathcal{T}_{\mu\nu}|0\rangle =0\quad\text{for elementary fields}\: ,\label{osdkjgfjd}
\end{align} 
to remove the inconsistency between \eqref{kjsdngnsdsga} and \eqref{kjsdngnsdsgb}. This is achieved by the law of nature \eqref{kjsdjmnsnsdg}, but not by \eqref{kdkjgdjnfdng}. 

Alternatively, the inconsistency between \eqref{kjsdngnsdsga} and \eqref{kjsdngnsdsgb} could of course be avoided if the assumption of relativistic covariance of the vacuum would be skipped. Then \eqref{iagnsaygrb} would not be valid, and \eqref{jmndnfgns} could be replaced by some non\bz{-}covariant method of regularization, like cut\bz{-}off regularization, which would avoid the change of signs. This possible scenario has been considered by Nikoli\'{c}\cite[sec.\,6]{Nikolic:prefFrame}: We could for example model the vacuum as filled with particles, like Dirac did in his hole theory\!\cite{Dirac:hole}. In this non\bz{-}covariant vacuum there would exist a preferred reference frame, in which the mean velocity of the vacuum particles is zero. Nikoli\'{c} noted that this assumption may seem less strange, if we remember that in well\bz{-}established cosmological models there exist preferred local reference frames, \ggie  the reference frames attached to the co\bz{-}moving observers in the FLRW metric\!\cite{Bartelmann:cosmology}. Obviously this method for the elimination of the inconsistency between \eqref{kjsdngnsdsga} and \eqref{kjsdngnsdsgb} would require a much more radical intervention into the foundations of quantum field theory than the introduction of the law of nature \eqref{kjsdjmnsnsdg}, and therefore seems not really attractive. 

\section{Discussion}\label{sec:discussion} 
The first \textsc{pro} argument asked: Why should different laws of nature apply to elementary fields versus discrete fields? And the first \textsc{contra} argument answered: Because the indeterminacy relations enforce zero\bz{-}point oscillations in case of discrete fields, but not in case of elementary fields, which are substrate\bz{-}less fields. It is probably fair to say that this \textsc{contra} argument is minimum as strong as the first \textsc{pro} argument. 

How should the second \textsc{pro} argument (the Casimir effect) be rated? In my point of view Casimir's heuristic method may be acknowledged as an ingenious alternative algorithm for the computation of van\bz{\,}der\bz{\,}Waals interactions, which has been applied --- and still is being applied by today --- with remarkable success to a large variety of systems. As an argument for the reality of an alleged zero\bz{-}point energy of the electromagnetic field, however, it is worth nothing. This judgment may sound harsh, but I think it is fully justified by the reasons explicated in section\;\ref{sec:caseff}\,. 

The suggested law of nature \eqref{kjsdjmnsnsdg} is not compatible with phase changes of elementary fields. The third and fourth \textsc{pro} arguments claim, however, that a phase change of the Higgs field and a phase change of an inflaton field actually have happened in the history of the universe. With regard to the electro\bz{-}weak phase change, I have argued in section\;\ref{sec:higgs} that the experimental confirmation of the Higgs Lagrangian \eqref{jsngnsdgbn} is not more a proof for the historic fact of an electro\bz{-}weak phase change, than the experimental confirmations of Maxwell's equations are a proof for the reality of the luminiferous ether. And the inflaton\bz{-}field\bz{-}phase\bz{-}change argument has been rated \al weak\ar  in section\;\ref{sec:inflation} because of our quite fragmentary knowledge of the physics prevailing so close to the big\bz{-}bang singularity of space\bz{-}time, and because the concept of phase changes of elementary fields seems fundamentally questionable. 

What about the second and third \textsc{contra} argument? It can hardly be denied that the power of each of them is impressing. Should we really seriously consider to sacrifice the Lorentz covariance of quantum field theory, as discussed at the end of section\;\ref{sec:signsincomp}? Only \eqref{kjsdjmnsnsdg}, but not \eqref{kdkjgdjnfdng}, is compatible with special relativity theory, as explicated in that section. And the power of the second \textsc{contra} argument (no gravitational effect of the alleged zero\bz{-}point energy) is literally obvious: We only need to open our eyes and look to the stars, to see that there is no zero\bz{-}point energy of elementary fields. 

Summing up: The first \textsc{pro} argument is easily balanced by the first \textsc{contra} argument. The second and the third \textsc{pro} argument are refuted. The fourth \textsc{pro} argument is weak. The second \textsc{contra} argument is \emph{very} strong. And the third \textsc{contra} argument as well is \emph{very} strong. It is probably no exaggeration to state that the arguments \textsc{contra} a zero\bz{-}point energy of elementary fields outweigh the \textsc{pro} arguments by many orders of magnitude. 

The ES\bz{-}tensor \eqref{kjsdjmnsnsdg} is 
\begin{ggitemize}
\ggitem{compatible with all observational evidence, and it is}
\ggitem{compatible with all established physical theories, in particular with SRT.}
\end{ggitemize}
No better justification could be imagined for a law of nature. The ES\bz{-}tensor \eqref{kdkjgdjnfdng}, on the other hand, is missing both criteria. 

Still many of us are reluctant to accept the different laws \eqref{kjsdjmnsnsdga} and \eqref{kjsdjmnsnsdgb} for the ES\bz{-}tensors of different types of fields, and hope for one universal law which would be valid for any type of field. The situation is reminiscent of the detection, that for fermion fields different quantization rules are valid than for boson fields: After Heisenberg\!\cite{Heisenberg:umdeutung} had due to ingenious guessing detected the basic rules of point\bz{-}particle quantum mechanics, Dirac\!\cite{Dirac:canquant} analyzed Heisenberg's strange non\bz{-}commutative matrix mechanics, and detected this general correlation between the observable quantities $A^\text{class},\ B^\text{class}$ of classical physics, and the non\bz{-}commutative algebra of their quantum\bz{-}mechanical counterparts $A^\text{qm},\ B^\text{qm}$\,: 
\begin{align}
A_s^\text{qm}B_t^\text{qm}-B_t^\text{qm}A_s^\text{qm}
&=i\hbar\big\{ A_s^\text{class}\; ,\: B_t^\text{class}\big\}\raisebox{-1.5ex}{\scriptsize Poisson\,brackets}\label{kjsngndfsg}\\ 
&=i\hbar\delta _{st}\hspace{4mm}\mbox{\parbox[t]{55mm}{if $B^\text{class}$ is the canonical conjugate\\ momentum of $A^\text{class}$}}\notag 
\end{align} 
While this method of canonical quantization could easily be generalized to boson fields\!\cite{Born:qm2,Dirac:quantemfeld}, it turned out that the minus sign in the commutator had to be changed to a plus sign in case of fermion fields\!\cite{Jordan:fermfeldquant}. 

With regard to the ES\bz{-}tensors of quantum fields, the situation is similar: The method \eqref{kdkjgdjnfdng} for the computation of ES\bz{-}tensors is a heritage of classical physics. The transfer of this formula to quantum theory resulted into ES\bz{-}tensors, which are obviously correct in case of quantum fields with material substrates. But it gives a nonsense\bz{-}result, \ggie  the huge vacuum energy \eqref{kmsmjgnsdffg}, in case of elementary fields, like Dirac's quantization algorithm \eqref{kjsngndfsg} gave correct results in case of boson fields, but nonsense results in case of fermion fields. 


In case of fermion versus boson fields, the community of physicists accepted different laws of nature. Why shouldn't we accept different laws of nature in case of elementary versus discrete quantum fields, even though this difference is backed by likewise strong theoretical and observational evidence? It is overdue that we now, after nine decades of hesitation, follow the 1928 suggestion of Jordan and Pauli, cited at the beginning of section\bz{\;}\ref{sec:gravitation}\,, \ggie  that we teach students and present in textbooks \eqref{kjsdjmnsnsdg}, but not \eqref{kdkjgdjnfdng}, as the correct law of nature.

\section*{Appendix A: Regularization \texorpdfstring{of $\boldsymbol{\langle 0|\,\mathcal{T}_{\mu\nu}\, |0\rangle}$}{}}
To \raisebox{4\baselineskip}[0pt][0pt]{\hypertarget{ta:appa}{}}regularize \eqref{pogfkgkjmgfj}, it's advantageous to switch to an infinite normalization volume, and apply spherical coordinates:\pagebreak[1] 
\begin{align}
\langle 0|\,\mathcal{T}_{\mu\nu}\, |0\rangle\!\ggstackrel[-.1]{\eqref{pogfkgkjmgfj}}\! F\hbar c\hspace{-.4em}\inte _0^{+\ggginfty }\hspace{-.4em}\frac{\dif\! k}{(2\pi )^3}\; 4\pi k^2\,\bigg( \eta _{\mu 0}\eta ^0{}_{\nu}\,\frac{k^2+K^2}{\sqrt{k^2+K^2}}\! -\!\frac{\eta _{\mu j}\eta ^j{}_{\nu}\, k^2}{3\sqrt{k^2+K^2}}\!\bigg)\notag\\ 
k&\equiv\sqrt{k_jk^j}\quad ,\quad K^2\equiv m^2c^2/\hbar ^2\label{sdmfmdfmhnmn}
\end{align} 
The finite $V_0$\bz{-}term could of course be skipped. 

\noindent{}Dimensional regularization
\ with $D=3-\epsilon $ and $0<\epsilon\in\mathbbm{R}$: 
\begin{align}
\langle 0|\,\mathcal{T}_{\mu\nu}\, |0\rangle &=\lim _{D\rightarrow 3}\frac{F\hbar c}{(2\pi )^D}\, \frac{2\pi ^{D/2}}{\Gamma (D/2)}\!\inte _0^{+\ggginfty }\!\dif\! k\, k^{D-1}\,\cdot\notag\\ 
&\qquad\cdot\bigg[ \frac{\eta _{\mu 0}\eta ^0{}_{\nu}3(k^2+K^2)-\eta _{\mu j}\eta ^j{}_{\nu}k^2}{3\sqrt{k^2+K^2}}\bigg]\notag  
\end{align} 
If 
\begin{subequations}\label{kdfjgjhxgf}\begin{align}
K\equiv mc /\hbar >0\ ,  
\end{align} 
then the substitution   
\begin{align} 
k=K\!\raisebox{-2.8mm}{\,\scalebox{.6}{$+$}\!}\sqrt{\frac{1-X}{X}}\quad\text{with}\ X\equiv\frac{K^2}{k^2+K^2} 
\end{align}\end{subequations} 
is possible, resulting into  
\begin{align}
\langle 0|\,\mathcal{T}_{\mu\nu}\, |0\rangle =&\lim _{\epsilon\rightarrow 0^+}\frac{F\hbar cK^{4}}{24\pi ^{3/2}\Gamma (3/2)}\cdot\Big(\frac{4\pi}{K^2}\Big) ^{\epsilon /2}\Gamma (-2+\epsilon /2)\,\cdot\notag\\ 
&\cdot\bigg[\frac{\eta _{\mu 0}\eta ^0{}_{\nu}3\Gamma (3/2)}{\Gamma (-1/2)}- 
\frac{\eta _{\mu j}\eta ^j{}_{\nu}\Gamma (5/2)}{\Gamma (+1/2)}\bigg]\ .   
\end{align} 
With the relations 
\begin{subequations}\begin{align} 
&\Gamma (\epsilon /2-n)=\Gamma (\epsilon /2)\cdot\prod _{\nu =1}^n\frac{1}{\epsilon /2-\nu}\quad ,\quad n=1,2,3,\dots\label{oadsgnhsgfrb}\displaybreak[1]\\ 
&\Gamma (\epsilon /2)=\frac{2}{\epsilon}-\gamma +\mathcal{O}(\epsilon ^2)\quad ,\quad\gamma\equiv 0.577215\dots\displaybreak[1]\\ 
&\Big(\frac{4\pi\, K_0^2}{K^2\, K_0^2}\Big) ^{\epsilon /2}\hspace{-.4em}=(K_0^2)^{-\epsilon /2}\Big[ 1+\frac{\epsilon}{2}\,\ln\Big(\frac{4\pi K_0^2}{K^2}\Big)\, +\mathcal{O}(\epsilon ^2)\Big] 
\end{align}\end{subequations}
one arrives at 
\begin{subequations}\label{ksdjmghnfdxyy}\begin{align} 
\langle 0|\,\mathcal{T}_{\mu\nu}\, |0\rangle &=-\eta _{\mu\nu}\,\frac{F\hbar cK^{4}}{32\pi ^{2}}\,\lim _{\epsilon\rightarrow 0^+}\Big[\frac{2}{\epsilon}-\gamma +\ln\Big(\frac{4\pi K_0^2}{K^2}\Big)\Big]\label{ksdjmghnfdxyya}\\ 
&=-\eta _{\mu\nu}\,\frac{F\hbar cK^{4}}{16\pi ^{2}}\lim _{w\rightarrow\ggginfty }\ln\!\Big(\frac{w}{K}\Big)\label{ksdjmghnfdx}\\ 
K&=mc/\hbar >0\ .\notag 
\end{align}\end{subequations} 
In the last step, modified minimal subtraction $\overline{\text{MS}}$ was applied. This result was first published by Akhmedov \cite{Akhmedov:vacener}.

Because of the substitution \eqref{kdfjgjhxgf}, the result \eqref{ksdjmghnfdxyy} is not valid for $K=mc/\hbar =0$. We can however consider the limit of arbitrary small --- and even unmeasurable small --- mass. Different from the assertions of \cite{Akhmedov:vacener,Martin:ccp}, the result \eqref{ksdjmghnfdxyy} stays infinite also in this case, for the following reason. One can find physical arguments to stop the limit ${B\rightarrow\gginfty }$ at some finite cut\bz{-}off\bz{-}wavenumber $B_{\text{max}}$ in cut\bz{-}off regularization, or to stop the limit ${\kappa\rightarrow\gginfty }$ at some finite invariant wavenumber $\kappa _{\text{max}}$ of the Pauli\bz{-}Villars counterterm (see appendix\;\hyperlink{ta:appb}{B}). But there is no physical argument known to stop $\epsilon\rightarrow 0$ (and consequently $w\rightarrow\gginfty  $) in dimensional regularization at any finite value $\epsilon _{\text{min}}$. Instead $\epsilon $ must become strictly zero, and thus the result \eqref{ksdjmghnfdxyy} is strictly infinite, even if a field's mass $m$ and invariant wavenumber $K$ should be unmeasurable small. 


Pauli-Villars regularization
: First the ES\bz{-}tensor's spatial components are regularized. The insertion of one counter\bz{-}term is sufficient. 
\begin{align}
\langle 0|\, \mathcal{T}_{il}\, |0\rangle\!\ggstackrel[-.1]{\eqref{sdmfmdfmhnmn}}\! 
-\eta _{ij}\eta ^j{}_{l}\frac{F\hbar c}{6\pi ^2}\!\inte _0^{+\ggginfty }\!\!\dif k\,\bigg( \frac{k^4}{\sqrt{k^2+K^2}}-\lim _{\kappa\rightarrow\ggginfty }\frac{k^4}{\sqrt{k^2+\kappa ^2}}\bigg)\! =\notag\displaybreak[1]\\ 
&=-\eta _{ij}\eta ^j{}_{l}\frac{F\hbar c}{16\pi ^2}\bigg(\! -K^4\,\ln\!\Big(\frac{K}{K_0}\Big) +\lim _{\kappa\rightarrow\ggginfty }\kappa ^4\,\ln\!\Big(\frac{\kappa}{K_0}\Big)\bigg)\notag\\ 
&=-\eta _{ij}\eta ^j{}_{l}\frac{F\hbar c}{16\pi ^2}\lim _{\kappa\rightarrow\ggginfty }\kappa ^4\,\ln\!\Big(\frac{\kappa}{K}\Big)\label{msngnsdfnhg}\\ 
K&\equiv mc/\hbar >0\ ,\ K_0\equiv (\text{wavenumber-unit})>0\notag
\end{align} 
Note that the finite term could be neglected. Furthermore in the last line $K_0$ has been chosen as $K$\,. 

As the Pauli\bz{-}Villars method is applicable only to integrals with negative powers of the invariant wavenumber $K$, \eqref{msngnsdfnhg} must be constrained to $m>0\, $, and the ES\bz{-}tensor's $00$\bz{-}component can not at all be regularized by that method. But as \eqref{msngnsdfnhg} is a correct result, it can be combined with the condition \eqref{iagnsaygrb}, and thus this general result can be concluded:  
\begin{align}
\langle 0|\,\mathcal{T}_{\mu\nu}\, |0\rangle\ggstackrel[.58]{\eqref{sdmfmdfmhnmn},\eqref{iagnsaygrb}}-\eta _{\mu\nu}\,\frac{F\hbar c}{16\pi ^2}\lim _{\kappa\rightarrow\ggginfty }\kappa ^4\,\ln\!\Big(\frac{\kappa}{K}\Big)\label{msngnsdfnhgx} \\ 
K&\equiv mc/\hbar >0\notag 
\end{align} 
The result is consistent with the result \eqref{ksdjmghnfdxyy} of dimensional regularization. With either regularization the results are infinite for both finite $m$ and $m\rightarrow 0\, $. 

\section*{Appendix B: Finite values\texorpdfstring{ for $\boldsymbol{\langle \,0\,|\,\mathcal{T}_{\mu\nu}\, |\,0\,\rangle}$}{}}
To \raisebox{4\baselineskip}[0pt][0pt]{\hypertarget{ta:appb}{}}extract finite values from the regularized result \eqref{msngnsdfnhgx}, we argue that the quantum field theories, which led to \eqref{msngnsdfnhgx}, are only low\bz{-}energy effective theories, and therefore the limit $\kappa\rightarrow\gginfty $ must be replaced by some finite wave\bz{-}number $\kappa _{\text{max}}$, like the integration over phonon wavenumbers is cut off at the maximum $|\boldsymbol{k}|$ of the first Brillouin zone. 

The invariant wave\bz{-}number $\kappa _{\text{max}}$ can hardly be less than about 
\begin{subequations}\begin{align}
\frac{10^{12}\text{eV}}{\hbar c}=5\cdot 10^{18}\text{m}^{-1}\ , 
\end{align} 
because otherwise some effect of $\kappa _{\text{max}}$ had been observed in high\bz{-}energy collider experiments. 

An upper limit can be estimated due to Heisenberg's indeterminacy relations: A particle cannot be localized with an accuracy better than half it's reduced Compton wavelength, \ggie  better than half it's inverted invariant wave\bz{-}number. This localization must not be less than two times the Schwarzschild radius $r_S\, $, because otherwise the particle would collapse to a black hole: 
\begin{align}
\Delta x&\approx\frac{1}{2\kappa}\geq 2r_S=\frac{4G}{c^2}\cdot\frac{\kappa\hbar}{c}\notag\\ 
\kappa &\leq\sqrt{\frac{c^3}{8G\hbar }}=\frac{1}{l_{\text{Planck}}\sqrt{8}}=2.2\cdot 10^{34}\text{m}^{-1}  
\end{align}\end{subequations} 
Thereby we get for the vacuum expectation value of the ES\bz{-}tensor of an elementary field the possible range 
\begin{align}
\langle 0|\,\mathcal{T}_{\mu\nu}\, |0\rangle\approx -\eta _{\mu\nu}F\cdot\Big(\, 5\cdot 10^{48}\dots\, 4\cdot 10^{111}\,\Big)\,\frac{\text{J}}{\text{m}^{3}}\ .\label{mksmgmnsdfg} 
\end{align}

\flushleft{\interlinepenalty=100000\bibliography{../gg}} 
\end{document}